\documentstyle{aipproc}

\begin{document}
\title{$B\to X_s\ell^+\ell^-$ in the vectorlike quark model}

\author{$^a$M. R. Ahmady\footnote{Email: mahmady@mta.ca}, $^b$M. Nagashima\footnote{Email: g0070508@edu.cc.ocha.ac.jp}, and $^b$A.
Sugamoto\footnote{Email:sugamoto@phys.ocha.ac.jp}}
\address{$^a$Department of Physics, Mount Allison University, Sackville, NB E4L 1E6 Canada\\
$b$Department of Physics, Ochanomizu University, 1-1 Otsuka 2, Bunkyo-ku, Tokyo 112, Japan}

\maketitle

\begin{abstract}
We extend the standard model by adding an extra generation of isosinglet up- and down-type quark pair which engage in weak interactions only via mixing with the three ordinary quark families.  It is shown that the generalized $4\times 4$ quark mixing matrix, which is necessarily nonunitary, leads to nonvanishing flavor changing neutral currents.  We then proceed to investigate various distributions and total branching ratio of the inclusive $B\to X_s\ell^+\ell^-$ ($\ell =e,\mu $) rare B decays in the context of this model.  It is shown that the shapes of the differential branching ratio and forward-backward asymmetry distribution are very sensitive to the value of the model parameters which are constrained by the experimental upper bound on $BR(B\to X_s\mu^+\mu^-)$.  We also indicate that, for certain values of the dileptonic invariant mass, CP asymmetries up to 10\% can be obtained. 

\end{abstract}
B factories, CLEO III and other dedicated B experiments are expected to observe new rare B decay channels and to improve the precision of those which have already been measured.  Radiative B decays, which proceed via loop effects, are quite crucial for observing the signals of new physics beyond the standard model (SM) in the near future.   One example of such models is  the extension of the SM to contain an extra generation of isosinglet quarks\cite{vqm}:  
\begin{equation}
\psi^i\equiv \left (\matrix{u^\alpha\cr d^\alpha}\right )\; ,\; \alpha=1...4 \; ,
\end{equation}
where
\begin{equation}
\left (\matrix{u^4\cr d^4}\right )\equiv \left (\matrix{U\cr D}\right ) \; .
\end{equation}
In other words, both left- and right-handed components of the additional pair of quarks, which are denoted $U$ and $D$ with charges $+2/3$ and $-1/3$, respectively, are $SU(2)_L$ singlets.  As a result, the Dirac mass terms of vectorlike quarks,i.e., 
\begin{equation}
m_U(\bar U_L U_R+\bar U_R U_L)+m_D(\bar D_L D_R+\bar D_R D_L) \;\; ,
\end{equation}
are allowed by electroweak gauge symmetry.  However, for ordinary quarks, gauge invariant Yukawa couplings to an isodoublet scalar Higgs field $\phi$,i.e.,
\begin{equation}
-f^{ij}_d{\bar\psi_L}^id_R^j\phi -f^{ij}_u{\bar\psi_L}^iu_R^j\tilde\phi \; +\; H.C. \;\; ,\; i=1,2,3\;,
\end{equation}
are responsible for the mass generation via spontaneous symmetry breaking.  The link between the vectorlike and the ordinary quarks are provided by the extra gauge invariant Yukawa couplings which can be written as follows:
\begin{equation}
-f^{i4}_d{\bar\psi_L}^iD_R\phi -f^{i4}_u{\bar\psi_L}^iU_R\tilde\phi \; +\; H.C. \;\; .
\end{equation}
After spontaneous symmetry breaking, Eqs. (4) and (5) along with Eq. (3) lead to $4\times 4$ mass matrices for the up- and down-type quarks:
\begin{equation}
{\bar d_L}^\alpha M_d^{\alpha\beta}d_R^\beta + {\bar u_L}^\alpha M_u^{\alpha\beta}u_R^\beta \; +\; H.C. \;\; ,
\end{equation}
In general, $M_d$ and $M_u$ are not diagonal, and to achieve diagonalization, unitary transformations from weak to mass eigenstates are necessary,i.e.,
\begin{equation}
u^\alpha_{L,R}={A^u_{L,R}}^{\alpha \beta}{u'_{L,R}}^\beta \; ,\;  d^\alpha_{L,R}={A^d_{L,R}}^{\alpha \beta}{d'_{L,R}}^\beta \;\; ,
\end{equation}
where we use prime to denote the mass eigenstates.  The interesting property of the vectorlike quark model (VQM) is that the above transformations result in the intergenerational mixing among quarks not only in the charged current sector but also in the neutral current interactions.  For example, the charged current interaction term which, when written in terms of weak eigenstates, involves only the three generations of ordinary quarks,i.e.,
\begin{equation}
{J^W_{CC}}^\mu=\sum^3_{i=1}I\frac{g}{\sqrt{2}}{\bar u_L}^i\gamma^\mu d_L^iW^+_\mu \; +\; H.C. \;\; ,
\end{equation}
transforms to
\begin{equation}
{J^W_{CC}}^\mu=\sum_{\alpha ,\beta =1}^4 I\frac{g}{\sqrt{2}}{\bar {u'}_L}^\alpha V^{\alpha \beta}\gamma^\mu {d'_L}^\beta W^+_\mu \; +\; H.C. \;\; ,
\end{equation}
where
\begin{equation}
V^{\alpha \beta}=\sum_{i=1}^3{({A^u_L}^\dagger)}^{\alpha i}{(A_L^d)}^{i\beta}\;\; ,
\end{equation}
when expressed in terms of the mass eigenstates.  It is straightforward to show that the $4\times 4$ quark mixing matrix $V$ is nonunitary
\begin{eqnarray}\nonumber
{(V^\dagger V)}^{\alpha \beta}&=&\delta^{\alpha\beta}-{({A_L^d}^{4\alpha})}^*{A_L^d}^{4\beta}\;\; ,\\
{(VV^\dagger )}^{\alpha\beta}&=&\delta^{\alpha\beta}-{({A_L^u}^{4\alpha})}^*{A_L^u}^{4\beta}\;\; . 
\end{eqnarray}
Therefore, a close examination of the neutral current sector reveals that flavor changing neutral currents (FCNC) like
\begin{equation}
 {J^Z_{NC}}^\mu =I\frac{g}{\cos\theta_w}\sum_{\alpha ,\beta =1}^4\left (I^q_wU^{\alpha\beta}{\bar{q'}_L}^\alpha\gamma^\mu {q'_L}^\beta -Q_q\sin^2\theta_w\delta^{\alpha\beta}{\bar{q'}}^\alpha\gamma^\mu {q'}^\beta \right ) \;\; ,
\end{equation}
where
\begin{equation}
U^{\alpha\beta}=\sum_{i=1}^3{({A_L^q}^{i\alpha})}^*{A_L^q}^{i\beta}=\delta^{\alpha\beta}-{({A_L^q}^{4\alpha})}^*{A_L^q}^{4\beta}=\left \{\matrix{
                       {(V^\dagger V)}^{\alpha\beta}\; ,\; q\equiv {\rm down-type} \cr
                       {(V V^\dagger)}^{\alpha\beta}\; ,\; q\equiv {\rm up-type}} \right.\;\; ,
\end{equation}
are consequently developed in the VQM.  It is exactly for this reason that rare B decays can serve as excellent venues to see the effects of the vectorlike quarks\cite{bvqm}.

The effective Lagrangian for $B\to X_s\ell^+\ell^-$ is the following
\begin{equation}
L_{\rm eff}=\frac{G_F}{\sqrt{2}}\left (A\bar sL_\mu b\bar\ell L^\mu\ell +B \bar sL_\mu b\bar\ell R^\mu\ell +2m_bC\bar sT_\mu b\bar\ell\gamma^\mu\ell \right )\;\; , 
\end{equation}
where
\begin{equation}
\nonumber L_\mu =\gamma_\mu (1-\gamma_5)\;\; , \;\; R_\mu =\gamma_\mu (1-\gamma_5)\;\; , 
\end{equation}
\begin{equation}
\nonumber T_\mu = i\sigma_{\mu\nu}(1+\gamma_5)q^\nu/q^2 \;\; .
\end{equation}
The coefficients $A$ and $B$ receive contributions from long-distance charm-quark loop ($c\bar c$ continuum) and the intermediate $\psi$ and $\psi'$ resonances as well as short-distance tree and one-loop diagrams (Fig.1).  However, C, the coefficient of the magnetic moment operator, is purely short-disctance. The details of the contributing terms to the effective Lagrangian can be found in Reference\cite{ans}. The total branching ratio of the dileptonic B decay is dominated by the resonance contributions.  However, by using cuts in the differential branching ratio around $\psi$ and $\psi'$ invariant mass, one can get information on the contributing short-distance physics\cite{ahmady96}.  

The VQM parameters which appear in our calculations are: the U-quark mass $m_U$, the nonunitarity parameter $U^{sb}=\vert U^{sb}\vert e^{i\theta}$, where $\theta$ is a weak phase, and $V^*_{4s}V_{4b}=U^{sb}-{(V^\dagger_{\rm CKM}V_{\rm CKM})}^{sb}$.  $V_{\rm CKM}$, which is the $3\times 3$ submatrix of the matrix $V$, consists of the elements representing mixing among three ordinary generations of quarks.  ${(V_{\rm CKM}^\dagger V_{\rm CKM})}^{sb}$ presents the deviation from unitarity of the 3-generation CKM mixing matrix in the VQM context.  The experimental upper bound $BR(B\to X_s\mu^+\mu^-)\le 5.8\times 10^{-5}$\cite{exp} is used, to put a rough constraint on the magnitude of $U^{sb}$,i.e., $\vert U^{sb}\vert\stackrel{<}{\sim}10^{-3}$.
Since the tree level diagram proportional to $U^{sb}$ (Fig. (1a)) contributes to the rare dileptonic decay channels, the one-loop terms with coefficient $V^*_{4s}V_{4b}$ are significant only if $V^*_{4s}V_{4b}\approx -{(V^\dagger_{\rm CKM}V_{\rm CKM})}^{sb}\gg U^{sb}$.  Thus, we parametrize our results in terms of ${(V^\dagger_{\rm CKM}V_{\rm CKM})}^{sb}/\vert V_{cb}\vert =\epsilon e^{i\phi}$ instead of $V^*_{4s}V_{4b}$, where $\epsilon =\vert {(V^\dagger_{\rm CKM}V_{\rm CKM})}^{sb}\vert /\vert V_{cb}\vert$ and $\phi$ is another weak phase of the model. 

We use $\vert V_{cs}\vert\approx 0.97$, $\vert V_{cb}\vert\approx 0.04$, $\vert V_{ts}\vert/\vert V_{cb}\vert\approx 1.1$, and $V_{us}^*V_{ub}\approx 0$\cite{pdg}, which are extracted from various experimental measurements and are not affected by the presence of the new physics.  $V_{cs}^*V_{cb}$ is taken to be real, as is the case, to a good accuracy, in the "standard" parametrization of the CKM matrix.  Consequently, $V_{ts}^*V_{tb}$, which is not known experimentally,  can be expressed in terms of the VQM parameters as:
\begin{equation}
\frac{V_{ts}^*V_{tb}}{\vert V_{cb}\vert}\approx \epsilon e^{i\phi}-\vert V_{cs}\vert \;\; .
\end{equation}

Figure 2 illustrates the differential branching ratio for some values of the VQM parameters as campared to the SM prediction.  We use $\epsilon =0.3$ and all constructive/destructive contribution possibilities of the extra, beyond the SM, terms are considered.  We observe that away from the resonances, where SD operators are dominant, the shift from the SM expectation, depending on the parameter values, can be quite significant.  To constrain the model parameters by using the experimental results on $BR(B\to X_s\mu^+\mu^-)$ reported in Ref. \cite{exp}, we calculate the total branching but excluding the resonances $\psi$ and $\psi'$ with a $\delta =\pm 0.1{\rm GeV}$ cut.  Our results are depicted in Fig. 3 in the form of  acceptable regions in the $\vert U^{sb}\vert$ versus $m_U$ plane for various choices of the relative sign of the extra contributions.  The most stringent constraint is obtained if the relative phases $\theta$ and $\phi$ both vanish (Fig. 3(a)).  

Figure 4 illustrates our results for the forward-backward asymmetry distribution of the decay $B\to X_s\mu^+\mu^-$ in the VQM as compared to the SM.  From Figs. 4(a) and 4(b) we observe that, when the tree level FCNC contributes constructively, even though the sign of the asymmetry remains the same as in the SM, its shape, away from the resonances, can be significantly different.  On the other hand, as is shown in Figs. 4(c) and 4(d), for the destructive contribution of the nonunitarity induced tree level term and large enough values of $\vert U^{sb}\vert$ or $m_U$, the sign of $A_{FB}$ can be opposite to what is predicted by the SM.  One important observable in this decay channel is the point of zero asymmetry in the forward-backward asymmetry distribution, which occurs somewhere below the resonance $\psi$.  Our investigation reveals that, as far as the VQM is concerned, the position of this point is quite stable and is not shifted very much from its SM value for various choices of the model parameters.

Since in the VQM, the effective Lagrangian contains terms with different CP-odd weak phases, as well as, LD continuum and resonance contributions which are sources of perturbative and nonperturbative CP-even strong phases, the direct CP asymmetry, unlike the SM, is expected to be non-zero.  The size of this asymmetry, which is zero for invariant dilepton masses below $2m_c$ threshold due to the vanishing strong phase, depends on the interplay of the various contributing terms.  Our results for certain choices of the model parameters are depicted in Fig. 5.  From Figs. 5(a) and 5(b), we observe that when the weak phase $\phi$ is large, smaller values of the nonunitarity parameter lead to significantly larger CP asymmetries sensitive to the $U$ quark mass, $m_U$.  In fact, for a negative relative sign of the tree level contribution ($\theta =\pi$) and $m_U=400 {\rm GeV}$, as is shown in Fig. 5(b), asymmetries of the order of 10\% can be achieved.

In conclusion, we have investigated various observables of the dileptonic rare B decay $B\to X_s\ell^+\ell^-$ in the presence of an extra generation of vectorlike quarks.  The measurement of this decay channel in the near future should provide more stringent constraints on the model parameters.   
 
{\bf\large Acknowledgement}\\
Special thanks to the Natural Sciences and Engineering research Council of Canada for support under the International Opportunity Fund grant number 237905-00.

\end{document}